\documentstyle[12pt, epsfig]{article}
\newcommand{\bea}   {\begin{eqnarray}}
\newcommand{\eea}   {\end{eqnarray}}
\begin{document}
\renewcommand{\thefootnote}{\fnsymbol{footnote}}

\thispagestyle{empty}

\title{On the Irreps of the $N$-Extended Supersymmetric Quantum Mechanics and Their Fusion Graphs.}
\author{Francesco Toppan${}^{a}$\thanks{{\em 
Talk given at the $22^{nd}$ Max Born Symposium in honour of the
$70^{th}$ birthday of Prof. J. Lukierski, Wroc\l aw, Sept. $2006$.
}}
\\ \\
${}^a${\it CBPF, Rua Dr.}
{\it Xavier Sigaud 150,}
 \\ {\it cep 22290-180, Rio de Janeiro (RJ), Brazil}\\
 {\it E-mail: toppan@cbpf.br}}
\maketitle
\begin{abstract}
In this talk we review the classification of the irreducible representations of the
algebra of the $N$-extended one-dimensional supersymmetric quantum mechanics presented in
hep-th/0511274.
We answer some issues raised in hep-th/0611060, proving the agreement of the results here contained 
with those in hep-th/0511274. 
We further show that the fusion algebra of the $1D$ $N$-extended supersymmetric vacua
introduced in hep-th/0511274 admits a graphical presentation. The $N=2$  
graphs are here explicitly presented for the first time.
\end{abstract}

\vfill
\rightline{CBPF-NF-019/06}

\newpage
\section{Introduction}

The one-dimensional $N$-extended supersymmetric quantum mechanics is an important and active research subject
in many areas of both mathematics and physics. We refer to \cite{Bellucci:2006} and \cite{Toppan:2006} for more comprehensive
recent reviews of some of its aspects which, due to space-time limitations, cannot be covered in the present talk.
Here we will focus on two main topics. At first, we make a review of the current status of the classification
of the irreducible linear representations of the algebra associated with the one-dimensional supersymmetric quantum mechanics.
In the following it will be presented a graphical interpretation of the $1D$ fusion algebra introduced in \cite{Kuznetsova:2005}.\par
Concerning the classification of the irreps, the basic references are \cite{Pashnev:2001} and \cite{Kuznetsova:2005}. In \cite{Pashnev:2001} it was
proven, essentially, that the irreps fall into classes of equivalence characterized by an associated Clifford algebra
irrep (the connection between Clifford algebras and extended supersymmetries of $1D$ quantum mechanics was previously shown
in \cite{{de Crombrugghe:1983},{Baake:1985},{Gates:1995}}). In \cite{Kuznetsova:2005} this result was used as the starting point to produce a classification of the irreps.
In this work a presentation of the results of \cite{Kuznetsova:2005} will be made. Some issues raised in \cite{Doran:2006bb} (see also \cite{Doran:2006aa}) will be answered, proving the compatibility of their results with
the \cite{Kuznetsova:2005} classification.
\par
It seems appropriate to present this seminar in a Symposium in honour of Jerzy Lukierski.
Even if our collaboration did not involve the topics here discussed, the results here presented, however, were made possible by applying a formalism first elaborated in our common works (especially \cite{Lukierski:2002}).
  
\section{The classification of the irreps}

The finite linear irreps of the $D=1$ $N$-extended supersymmetry algebra
\begin{eqnarray}\label{susyalg}
\{ Q_i,Q_j\}&=& \delta_{ij} H
\end{eqnarray}
(where the $Q_i$'s, $i,j=1,\ldots , N$, are the supersymmetry generators and $H$ is the hamiltonian)
are expressed by the set of $(n_1,n_2,\ldots, n_l)$ symbols representing the
field-contents of the irreps. The non-negative integers $n_i$'s specify the number
of fields of dimension $d_i=d_1+\frac{i-1}{2}$ entering an irrep (the constant $d_1$
can be arbitrarily chosen). The fields whose dimension differs by $\frac{1}{2}$ have opposite statistics 
(bosonic/fermionic). The number $l$ specifies the number of different dimensions
of the fields entering an irrep and is referred to as the ``length" of the irrep; $l$ must satisfy the
condition $l\geq 2$, with the $l=2$ irreps being known in the literature as the ``minimal-length" or ``root"
multiplets.  In \cite{Kuznetsova:2005} 
it was explicily presented the complete list of the allowed $(n_1,n_2,\ldots, n_l)$ field contents for $N\leq	10$. An algorithmic construction for computing the field contents for
arbitrarily large values of $N$ was produced and selected $N>10$ examples were given. The list in \cite{Kuznetsova:2005} is understood as follows: for any $N$, $(n_1,n_2,\ldots, n_l)$
is present if and only if there exists at least one $N$-irrep with the given field content.
As an example, the length-$4$ $(1,7,7,1)$ field content is present for $N=5,6,7$, but not
for $N=8$, meaning that there are no irrep with the given field content for $N=8$, but there
is (at least one) such irrep for $N=5,6,7$. \par
The construction of \cite{Kuznetsova:2005} heavily relied on the \cite{Pashnev:2001} results
which we briefly summarize here. All $N$-irreps of length $l\geq 3$ are obtained from the set
of ${\overline Q}_i$ operators acting on root multiplets after applying an {\em acceptable}
dressing transformation ${\overline Q}_i\rightarrow D{\overline Q}_i D^{-1}= {Q}_i' $. The dressing
operator $D$ is a diagonal operator whose entries are either $1$ or positive powers
of $H$. ``Acceptable" refers to the fact that the whole set of ${Q}_i'$ transformed operators
must be regular (that is, as matrix operators, they must not contain any entry with $\frac{1}{H}$ or higher poles). Two distinct acceptable operators $D_1$, $D_2$ leading to the same field content applied on the
same set of ${\overline Q}_i$ root multiplets operators are given by a permutation of their diagonal
entries. $D_1, D_2$ are obviously related by a similarity transformation,
$D_2=S D_1S^{-1}$ (it is worth recalling that the exchange of the diagonal elements in,
e.g., a $2\times 2$
diagonal matrix $D$ is recovered in terms of the $2\times 2$ similarity matrix $S=\left( 
\begin{array}{cc}
$0$ &$1$\\
$1$ &$0$
\end{array}
\right)
$).
 Similarity transformations between two acceptable dressings of given
field-content and for a fixed set of ${\overline Q}_i$ operators acting on root multiplets form a group of transformations which corresponds to a subgroup ${\widetilde G}$ of the permutation group of the diagonal
elements of the dressing transformations. \par
Concerning the length-$2$ root multiplets the situation is the following. They are recovered
by an associated Clifford irrep of Weyl type (i.e., whose generators are in block antidiagonal
form) through the following position
\begin{eqnarray}\label{length2irrep}
{\overline Q}_i&=& \frac{1}{\sqrt 2}\left( \begin{tabular}{cc} $0$& $\sigma_i$\\
${\widetilde \sigma}_i\cdot H$& $0$
\end{tabular}
\right) ,
\end{eqnarray}
where $\sigma_i$ and ${\widetilde\sigma}_i$ are matrices entering the associated Clifford generators
\begin{eqnarray}\label{weylclifford}
\Gamma_i =\left( \begin{tabular}{cc} $0$& $\sigma_i$\\
${\widetilde \sigma}_i$& $0$
\end{tabular}
\right)\quad &,&\quad\{ \Gamma_i,\Gamma_j\}= 2\delta_{ij}.
\end{eqnarray}
The $N$ Clifford generators entering (\ref{weylclifford}) are recovered from
the block-antidiagonal space-like generators of the $Cl(p,q)$ Clifford algebras (with $(p,q)$ signature) according to the following scheme:
{{
{ {{\begin{eqnarray}&\label{oxidation1}
\begin{tabular}{cllll}
$Cl(2+8m,1)  $&$\rightarrow$&$N=1$ & mod &$ 8$\\
$Cl(3+8m,2)  $&$\rightarrow$&$N=2$ &mod &$ 8$\\ 
$Cl(4+8m,3)  $&$\rightarrow$&$N=3$& mod &$8$\\ 
$Cl(5+8m,0)   $&$\rightarrow$&$N=3, 4$& mod &$ 8$\\ 
$Cl(6+8m,1)  $&$\rightarrow$&$N=5$& mod &$8$\\ 
$Cl(9+8m,0)  $&$\rightarrow$&$N=5,6,7, 8$& mod &$ 8$\\
\end{tabular}&\end{eqnarray}}} }  
}} 
The maximal value for $N$ corresponds to $N=p-1$ (the ``oxidized" cases \cite{Kuznetsova:2005}). The reduced 
supersymmetries (for $N<p-1$) are obtained by selecting a proper subset of the 
block antidiagonal $Cl(p,q)$ space-like generators. Notice that, unlike the other values
of $N$, the $N=3,5$ mod $8$ cases can be recovered in two different ways.\par
For a fixed value of $N$, the $N$ Clifford generators entering (\ref{weylclifford}) can be
uniquely chosen up to similarity transformations (this result is in consequence of the
unicity of the real irreducible Clifford algebra representations for $p-q\neq 1,5$ mod $8$).\footnote{Please notice that this new set of similarity transformations acts on supersymmetry
operators for root multiplets; it should not be confused with the set of similarity transformations
acting on dressing operators.}
This important property applies in particular to the reduced values of $N$, implying that
two different choices of the 
$N<p-1$ proper subset of block antidiagonal space-like generators are equivalent. 
It also implies that the two ways in (\ref{oxidation1}) of recovering the $N=3,5$ mod $8$ supersymmetries are equivalent, producing root multiplets which are related by similarity
transformations. 
For any given $N$ the (\ref{weylclifford}) Clifford generators associated to supersymmetric
root multiplets can be canonically chosen. They can be presented as matrices whose non-vanishing
entries are $\pm 1$. A group $G$ of similarity transformations relates all choices of Clifford generators 
whose non-vanishing entries are $\pm 1$. The dressing transformations, applied to each set of Clifford generators of this type, produce irreps with the same field-contents. 
Taking into account these properties, the \cite{Kuznetsova:2005} classification of the field-contents produces a classification of the linear finite irreps of the $D=1$ $N$-extended
supersymmetry. The $(n_1,\ldots, n_l)$
symbol uniquely characterizes the irreps upon which the $D{\overline Q}_i D^{-1}$ supersymmetry operators act. The ${\overline Q}_i$ operators acting on root multiplets are related by the group $G$ of similarity
transformations, while the acceptable dressing operators $D$ are related by the group ${\widetilde G}$ of similarity transformations. Under this equivalence class of transformations, $(n_1,\ldots , n_l)$ uniquely specifies an irrep. \par
The complete list of $(n_1,\ldots, n_l)$ irreps for $N\leq 10$ is furnished
in \cite{Kuznetsova:2005} and will not be reproduced here.

\section{Irreps fusion algebras and the associated graphs}

The notion of fusion algebra of the supersymmetric vacua of the $N$-extended one dimensional
supersymmetry was introduced in \cite{Kuznetsova:2005}. 
The tensoring of two zero-energy vacuum-state irreps (irreps associated with the zero
energy eigenvalue of the hamiltonian operator $H$) can be symbolically written as
\
\begin{eqnarray}\label{fusion}
\relax [i]\times [j] &=& {N_{ij}}^k [k]
\end{eqnarray}
where
${N_{ij}}^k$ are non-negative integers specifying the decomposition of the tensored-products irreps into its irreducible constituents. The ${N_{ij}}^k$ integers satisfy a fusion algebra 
with the following properties\par
$1$) Constraint on the total number of component fields,
\begin{eqnarray}\label{ineqconstr}
\forall ~ i,j\quad \sum_k {N_{ij}}^k &=&2d
\end{eqnarray} 
where $d$ is the number of bosonic (fermionic) fields
in the given irreps.\par
$2$) The symmetry property 
\begin{eqnarray}\label{fusionsymm}
{N_{ij}}^k&=&{N_{ji}}^k
\end{eqnarray}

$3$) The associativity condition,
\begin{eqnarray}\label{assoc}
\relax [i]\times ([j]\times[k]) &=&([i]\times[j])\times[k]
\end{eqnarray}
which implies the commutativity of the  $(N_i)_j^k\equiv N_{ij}^k$ fusion matrices.
\par
Fusion algebras can also be nicely presented in terms of their associated graphs. The $N=1$ and $N=2$ fusion graphs are produced in the Appendix (with two subcases for each $N$, according to whether or not the irreps are distinguished w.r.t. their bosonic/fermionic statistics). 
Let us discuss here how to present the \cite{Kuznetsova:2005} results in graphical form. The irreps correspond to points.
$N_{ij}^k$ oriented lines (with arrows) connect the $[j]$ and the $[k]$ irrep if the decomposition (\ref{fusion}) holds.
The arrows are dropped from the lines if the $[j]$ and $[k]$ irreps can be interchanged.  The $[i]$ irrep should
correspond to a generator of the fusion algebra. This means that the whole set of ${N_l=N_{lj}}^k$ fusion matrices
is produced as sum of powers of the ${N_i=N_{ij}}^k$ fusion matrix.\par
Let us discuss explicitly the $N=2$ case. We obtain the following list of four irreps (if we discriminate their statistics):
\begin{eqnarray}&
\relax[1]\equiv (2,2)_{Bos};~
\relax [2]\equiv (1,2,1)_{Bos}
;~
\relax[3]\equiv (2,2)_{Fer};~
\relax [4]\equiv (1,2,1)_{Fer}&
\end{eqnarray}
The corresponding $N=2$ fusion algebra is realized in terms of four $4\times 4$, mutually commuting, matrices given by
\small{
\begin{eqnarray}&
N_1 = \left(\begin{array}{cccc} $1$&$2$&$1$&$0$\\
$0$&$2$&$0$&$2$\\
$1$&$0$&$1$&$2$\\
$0$&$2$&$0$&$2$
\end{array}\right)\equiv X; 
N_2 = N_4 =\left(\begin{array}{cccc} $0$&$2$&$0$&$2$\\
$0$&$2$&$0$&$2$\\
$0$&$2$&$0$&$2$\\
$0$&$2$&$0$&$2$
\end{array}\right)\equiv Y;
N_3 = \left(\begin{array}{cccc} $1$&$0$&$1$&$2$\\
$0$&$2$&$0$&$2$\\
$1$&$2$&$1$&$0$\\
$0$&$2$&$0$&$2$
\end{array}\right)\equiv Z.&\nonumber\\
&&\label{XYZ}
\end{eqnarray}}
The fusion algebra admits three distinct elements, $X,Y,Z$ and one generator (we can choose either $X$ or $Z$), due to the relations
\begin{eqnarray}
Y=\frac{1}{8}(X^3-2X) &,& Z=-\frac{1}{4}(X^3-6X^2+4X).
\end{eqnarray}
The vector space spanned by $X,Y,Z$ is closed under multiplication
\begin{eqnarray}
X^2=Z^2=ZX&=& X+2Y+Z,\nonumber\\
XY=Y^2=YZ&=&4Y.
\end{eqnarray}
This fusion algebra corresponds to the ``smiling face" graph (Figure $4$ in the Appendix).

\section{Conclusions}

The supersymmetric quantum mechanism is a fascinating subject with several open problems.
The potentially most interesting one concerns perhaps the construction of off-shell invariant
actions with the dimension of a kinetic term for large values of $N$ (let's say $N>8$).
They could provide some hints towards an off-shell formulation of higher-dimensional
supergravity and $M$-theory. The fusion algebras, which encode the information of the
decomposition of tensor representations, could provide useful in attacking this problem.
\par
Concerning the representation theory itself, some questions are still opened.
The authors of \cite{Doran:2006bb} pointed out the existence of inequivalent 
(starting from $N\geq 5$) supersymmetry irreps with the same field content. They explicitly
discussed the $N=5$ $(6,8,2)$ and the $N=6$ $(6,8,2)$ irreps, producing in both cases
two inequivalent irreps. These results are in agreement
with those in \cite{Kuznetsova:2005}. At first it must be noticed that $(6,8,2)$ is an 
admissible field content for both $N=5$ and $N=6$ irreps, see \cite{Pashnev:2001}. 
The inequivalences obtained in \cite{Doran:2006bb}
correspond to a different notion of the equivalence relation than the one here discussed
(their equivalence class is w.r.t. the general linear transformations
of the supersymmetry generators and/or fields). It produces a refinement of the 
equivalence relation here employed.  
To spot the differences, one can use the valid analogy of the classification
of simple Lie algebras. Simple Lie algebras over the complex numbers are classified by the
Dynkin's diagrams, while simple Lie algebras over the reals are obtained by the real forms.
The $(n_1,\ldots, n_l)$ field contents work as Dynkin's diagrams, uniquely
specifying the irreps under the class of equivalence here discussed. \par
Concerning the classification of irreps, the present status is the following.
The complete classification of the irreps under the equivalence relation 
here discussed was produced in \cite{Kuznetsova:2005} (explicit results were furnished for
$N\leq 10$).  At present, no classification of irreps is yet available under the \cite{Doran:2006bb} notion of the equivalence relation. \par
{~}\\

{\large{\bf  Appendix: the $N=1,2$ fusion graphs.}}
{~}\\
{~}\\
{~}
We present here four fusion graphs of the $N=1$ and $N=2$ supersymmetric
quantum mechanics irreps. The ``$A$" cases below correspond to ignore the statistics
(bosonic/fermionic) of the given irreps. In the ``$B$" cases, the number of fundamental irreps
is doubled w.r.t. the previous ones, in order to take the statistics of the irreps into account. 
The construction of the graphs is discussed in the main text.
\\
\begin{figure}[htbp]
\epsfig{file=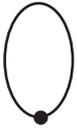} 
\vspace{-2.2 cm}
\vspace{3cm}
\caption{Fusion graph of the N=1 superalgebra, $A$ case, $1$ irrep ($(1,1)$), no bosons/fermions distinction.}
\end{figure}

\vspace{1cm}
\begin{figure}[htbp]
\epsfig{file=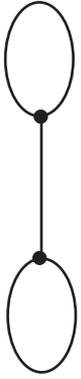} 
\vspace{-2.2 cm}
\vspace{3cm}
\caption{Fusion graph of the N=1 superalgebra, $B$ case, $2$ irreps ($(1,1)_{Bos}$ and $(1,1)_{Fer}$) with bosons/fermions distinction.}
\end{figure}

\vspace{1cm}
\begin{figure}[htbp]
\epsfig{file=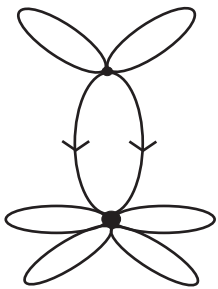} 
\vspace{-2.2 cm}
\vspace{3 cm}
\caption{Fusion graph of the N=2 superalgebra, $A$ case, $2$ irreps ($(2,2)$ and $(1,2,1)$),
no bosons/fermions distinction.}
\end{figure}

\vspace{1cm}
\begin{figure}[htbp]
\epsfig{file=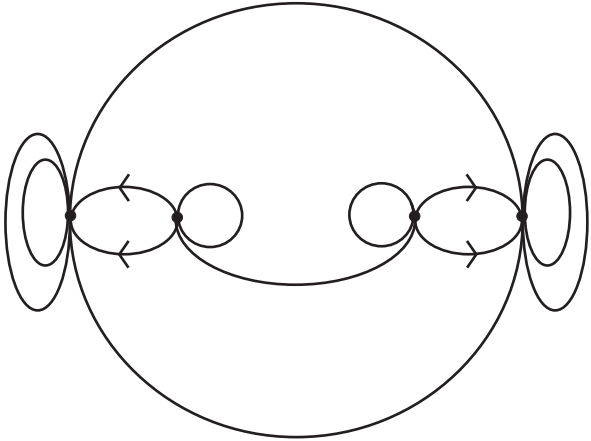} 
\vspace{-2.2 cm}
\vspace{3 cm}
\caption{Fusion graph of the N=2 superalgebra, $B$ case, $4$ irreps, bosons/fermions distinction,
``the smiling face". From left to right the four points correspond to the $[2]-[1]-[3]-[4]$ irreps, respectively. 
The lines are generated by the $N_1\equiv X$ fusion matrix, see (\ref{XYZ}).}
\end{figure}

\newpage
{\large{\bf Acknowledgments}}
{~}\\{~}\\
I am grateful to the organizers of the Max Born Symposium for the kind invitation. I am pleased to thank my collaborators
Z. Kuznetsova and M. Rojas for the main results here presented. Concerning fusion graphs, I have profited of helpful discussions with R. Coquereaux, while F.V. Fortaleza de Vasconcelos is credited for the drawings.
This work received financial support from CNPq and FAPERJ.

\end{document}